# Comparative study on high temperature mechanical behavior in 3YTZP containing SWCNTs or MWCNTs

Miguel Castillo-Rodríguez[1*], Antonio Muñoz[2] and Arturo Domínguez-Rodríguez[2]

[1]Instituto de Ciencia de Materiales de Sevilla, CSIC-Universidad de Sevilla, Avda. Américo Vespucio 49, 41092 Sevilla, Spain

[2]Departamento de Física de la Materia Condensada, Facultad de Física, Universidad de Sevilla, Apartado 1065, 41080 Sevilla, Spain

**Abstract:** Effects on mechanical properties of the presence of either single-walled carbon nanotubes (SWCNTs) or multi-walled carbon nanotubes (MWCNTs) in a 3YTZP matrix have been investigated in this work. Thus, monolithic 3YTZP and 3YTZP containing 2.5 vol% either SWCNTs or MWCNTs were fabricated by Spark Plasma Sintering (SPS) at 1250 ºC. Samples were crept at temperatures between 1100 and 1200 ºC and stresses between 5 and 230 MPa. Raman spectroscopy measurements indicate the absence of severe damages in the CNTs structure after sintering and testing. Scanning electron microscopy studies show that microstructures do not evolve during creep tests. Mechanical results point out that monolithic 3YTZP exhibits a higher creep resistance than composites since CNTs facilitate grain boundary sliding during high-temperature deformation. SWCNTs and MWCNTs have a similar effect on the high temperature mechanical behavior in 3YTZP where the bundle length and the level of dispersion of CNTs play a crucial role.

**Keywords:** Nanotubes; Zirconia; Composites; Creep; Mechanical properties;

*Corresponding author: miguelcr@us.es. Instituto de Ciencia de Materiales de Sevilla, CSIC-Universidad de Sevilla, Avda. Américo Vespucio 49, 41092 Sevilla, Spain.   Tel: Int-34-95 455 09 64, Fax: Int-34-95 461 20 97.

# 1. Introduction

Ceramic materials possess an extraordinary combination of remarkable physical properties (high strength, hardness, thermal-chemical stability and wear resistance) and therefore they are widely used for technological applications. However, ceramic brittleness still remains a major concern since it restricts their use as structural materials [1]. CNTs possess extraordinary mechanical, electrical and thermal properties [2-4] and therefore since their discovery the ceramic materials community quickly considered CNTs as one of the most promising candidates for improving ceramic materials toughness. Thus, many efforts have been performed in order to sinter ceramic/CNTs composites with enhanced mechanical properties.

Nonetheless, there is still a big controversy whether the incorporation of CNTs in a ceramic matrix produces a strengthening and toughening effect or not [5-11]. Regarding room temperature mechanical properties, the technique used to measure the fracture toughness, Vickers indentation fracture (VIF) or single edge notched beam (SENB), seems to contribute to this controversy. Thus, although VIF method has been the most widely used, some authors assert that SENB provides more reliable and lower toughness values than VIF tests [5,6,8]. This fact is clearly observed in $Al_2O_3$/SWNTs composites since Vickers indentation gives higher fracture toughness values than those measured in fully dense pure $Al_2O_3$ [7], however no increase is found from SENB tests [8]. Together with this discrepancy on the fracture toughness measurement method, it seems that there is a different toughening effect depending on the CNTs type, SWCNTs or MWCNTs, added to the ceramic matrix. This is observed in

3YTZP/CNTs composites. Mazaheri *et al*. [12] reported that fracture toughness, measured by VIF, improved by a factor of two in 3YTZP containing 0.5-5 wt% MWCNTs compared to monolithic 3YTZP, whereas Poyato *et al*. [10] found a decrease in hardness and fracture toughness when increasing the SWCNTs content up to 10 vol% in 3YTZP/SWCNTs composites.

The incorporation of SWCNTs or MWCNTs to a 3YTZP ceramic matrix may also influence differently its high temperature mechanical behaviour. Castillo-Rodríguez *et al*. [13] reported that the incorporation of 2.5 vol% SWCNTs in a 3YTZP ceramic matrix has a softening effect; since 3YTZP/SWCNTs composites exhibit at 1200 ºC a creep rate about 60 times higher than that corresponding to the monolithic 3YTZP. They argued that SWCNTs makes GBS easier during deformation instead of acting as a reinforcing element. The opposite effect was found by Mazaheri *et al*. [14] in zirconia containing different quantities of MWCNTs instead of SWCNTs, and they reported a creep rate decrease of three orders of magnitude for a composite with 5 wt% of MWCNTs concentration compared to monolithic zirconia. This different influence depending on the CNTs type has been also observed in $Al_2O_3$/CNTs composites. Zapata *et al*. [15] reported a strain rate in 10 vol% SWNT-reinforced $Al_2O_3$ of two orders of magnitude lower than in pure alumina of the same grain size. On the contrary, Estili *et al*. [16] observed a much lower flow stress in 20 vol% surface acid-treated $Al_2O_3$/MWCNTs composites than in pure alumina. Moreover, Huang *et al*. [17] observed that plastically un-deformable alumina ceramic showed promising superplastic behaviour even at temperatures as low as 1300 ºC by the addition of 0.5 wt% of BN nanotubes.

All these results evidence that the nature of nanotubes plays a crucial role in the final mechanical properties of the ceramic/CNTs composites. In this work we have sintered 3YTZP containing either SWCNTs or MWCNTs in order to investigate their influence on the high temperature mechanical behaviour. For the sake of comparison, monolithic 3YTZP has been also fabricated using the same sintering conditions. Creep experiments at temperatures between 1100 and 1200 ºC have been conducted in all fabricated specimens. Experimental results of monolithic 3YTZP and 3YTZP/CNTs composites have been analysed and compared, with the advantage of having fabricated all samples using the same sintering conditions. Thus, the effects of CNTs (SWCNTs or MWCNTs) on microstructure and mechanical properties have been more clearly identified.

## 2. Experimental Procedure

*2.1. Starting materials, powder processing and sintering*

3 mol% yttria stabilized zirconia 3YTZP powder (Nanostructured and Amorphous Materials Inc., Houston, TX), with 40 nm average particle size and 99 % purity, was annealed in air at 1250 ºC for one hour and subsequently ball milled for 3 hours with a frequency of 25 vibrations/s (Model MM200, Retsch GmbH, Haan, Germany).

Regarding CNTs, two types have been used in this work. On the one hand, commercially available 90 % purified SWNTs, with a typical bundle length of 0.5-1.5 μm and diameter between 4-5 nm, were provided by Carbon Solutions Inc. (Riverside, CA). They were COOH-functionalized following the same routine than Poyato *et al.* [10], although at the end of the routine the acid-treated SWCNTs were freeze-dried instead of being dried on hot plate. On the other hand, it has been used commercial COOH-functionalized MWCNTs (purity > 95 %), with 15±5 nm in diameter and between 1 and 5 μm in length.

The powder processing routine has been the same than for composites called "1bFD-probe"in Ref [13], which generated the optimal CNTs dispersion throughout the 3YTZP matrix. In this work 3YTZP/CNTs composites have been fabricated containing 2.5 vol% either SWCNTs or MWCNTs. Basically, the powder processing routine consisted on: Aqueous colloidal processing ACP [18] where CNTs (SWCNTs or MWCNTs) and 3YTZP powder were sonicated separately in two high pH solutions (distilled water + $NH_3$ solution until a pH=12) for 1 hour in an ultrasonic probe (Model KT-600, Kontes Inc., Vineland, New Jersey). Later, they were suspended in the same high pH solution, and sonicated again for 30 min. After mixing 3YTZP powder and CNTs (SWCNTs or MWCNTs) in the high pH solution, they were frozen immediately by immersing the solution into liquid nitrogen in order to avoid the 3YTZP powder decantation. Finally the blend was freeze-dried.

3YTZP powders and composite powders were sintered by Spark Plasma Sintering (Model Dr Sinter 1050, Sumitomo Coal Mining Co. LTD, Tokyo, Japan) at 1250 ºC, with a constant uniaxial pressure of 75 MPa for 5 min and in ~ $10^{-2}$ mbar vacuum. The heating and cooling rates were 300 ºC/min and 50 ºC/min, respectively. Relative densities higher than 99 % were obtained (Table I). Densities of the sintered specimens were measured by Archimedes' method taking a density value of 1.3 g/cm$^3$ and 2.1 g/cm$^3$ for SWCNTs and MWCNTs respectively, as it was given by the providers.

2.2. Microstructural and mechanical characterization

The microstructure of the sintered samples was examined by light microscopy (Model Leica DMRE, Leica Microsystems GmbH, Germany) and by high resolution scanning electron microscopy HRSEM (Model HITACHI S5200, Hitachi High-Technologies Corporation, Tokyo, Japan) operating at 5kV. Prior to observation, the samples were ground and polished with diamond paste of grain sizes down to 1μm. Then, they were annealed for 15 min at 1100 ºC to reveal grain boundaries. To study the 3YTZP grain morphology, the equivalent planar diameter d = $(4 \times area/\pi)^{1/2}$, and the shape factor F=$4\pi \times area/(perimeter)^2$ were measured from HRSEM micrographs. In the case of 3YTZP/CNTs composites, these same parameters (D and F) were evaluated to characterize the morphology of CNTs agglomerates. Moreover the surface density of CNT agglomerates has been also, estimated from the area fraction covered by them in low magnification SEM micrographs and also by means of light microscope observations. Fracture cross-sections from composite specimens were also investigated by HRSEM, and by Raman spectroscopy to study CNTs integrity after sintering and after creep tests, using a dispersive microscope (Model LabRAm Horiba Jobin Yvon, Horiba Ltd, Kyoto, Japan). Measurements were done with a green laser (He–Ne 532.14 nm), 20-mW, 600 g/mm grating and without filter.

Samples were cut and ground as parallelepipeds of approximate dimensions 5 mm × 2.5 mm × 2.5 mm. Uniaxial compression creep tests were performed on a prototype creep machine [19] at temperatures between 1100 and 1200 ºC, in a controlled argon atmosphere to avoid the oxidation of the 3YTZP ceramic matrix and the combustion of CNTs, and stresses ranging between 5 and 230 MPa. The creep curves were analysed using the standard phenomenological creep equation:

$$\dot{\varepsilon} = A \frac{\sigma^n}{d^p} \exp\left(-\frac{Q}{KT}\right) \quad (1)$$

being $A$ a stress and temperature independent term reflecting the dependence of the strain rate on the microstructural features of the material (composition, amount and physical properties of glassy phases, grain morphology, etc.), $\sigma$ the applied stress, $d$ the grain size, $K$ the Boltzmann's constant, and T the absolute temperature. The parameters n, p, and Q (generically known as creep parameters) are, respectively, the stress and grain size exponents, and the apparent activation energy for creep.

3. Results and Discussion

Microstructure of samples is shown in Figure 1, where 3YTZP grains are faceted with sharp triple points. In case of 3YTZP/SWCNTs or 3YTZP/MWCNTs composites, CNTs are forming bundles surrounding 3YTZP grains (Figures 1b and 1c). They are also visible in deformed composites, so their integrities have been preserved during sintering and mechanical testing (Figures 1d and 1e). Table I shows microstructural characterization of the 3YTZP grains in monolithic and composites materials. In monolithic 3YTZP we have obtained an equivalent planar diameter d of about 0.27 µm, whereas slightly smaller values, d=0.20 µm and d=0.23 µm, are obtained for 3YTZP/SWCNTs and 3YTZP/MWCNTs composites respectively. This is pointing out that the presence of CNTs hinders grain growth during sintering, as it has been previously reported by other authors [20,21]. This hindering effect seems to be more effective in SWCNTs than in MWCNTs, probably due to the smaller bundle diameter of SWCNTs which allow them to fit better the 3YTZP grain boundaries and then impeding more effectively the grain growth during sintering. The shape factor $F$ obtained in all specimens is close to 0.7, indicating that grains are similarly equiaxed in monolithic 3YTZP and composites. These morphological parameters, $d$ and $F$, have been measured in deformed specimens obtaining similar values (Table I). These results indicate that microstructures do not evolve during creep tests. Regarding composites, CNTs agglomerates have been also characterized by measuring the average size $d$, the shape factor $F$ and the surface density of CNTs agglomerates $\rho_s$. Results are shown in Table I, and it is observed that 3YTZP/MWCNTs composites exhibit higher surface density and slightly bigger and more elongated CNTs agglomerates than 3YTZP/SWCNTs composites. Then SWCNTs seem to be better dispersed throughout the 3YTZP matrix than MWCNTs. The reason could come from differences in the functionalization process undergone by both types of CNTs, since SWCNTs were COOH functionalized following the same routine than Poyato *et al.* [10], and MWCNTs were provided already functionalized. Moreover, the larger bundle length and diameter of MWCNTs compared to SWCNTs could favor the formation of agglomerates. Anyway, morphological parameter values for SWCNTs and

MWCNTs are quite close to those exhibited by composites with optimal CNTs distribution processed by Castillo *et al.* [13] following the same routine (sample 1bFD-probe in Ref [13]). That means that good CNTs dispersions have been achieved for both composites in this work.

HRSEM observations (Figure 1) suggest that CNTs have kept their integrity after sintering and after the creep test. Raman spectroscopy has been performed in specimens prior to and after creep tests in order to confirm the absence of severe damages in SWCNTs (Figure 2a) or MWCNTs (Figure 2b) structures. For the sake of comparison, Raman spectra measured in monolithic 3YTZP and in SWCNTs or MWCNTs are also shown. Composites show on the one hand peaks at 165, 260, 320, 465, 610 and 643 cm$^{-1}$ which corresponds to the six Raman bands predicted for theoretical tetragonal zirconia. On the other hand, they also show the typical mode bands due to the presence of CNTs. In case of 3YTZP/SWCNTs composites, radial breathing mode bands located about 150-200 cm$^{-1}$, and the G mode corresponding to the tangential shear mode of carbon atoms (~1550-1600 cm$^{-1}$). Unsymmetrical G mode is characteristic of SWCNTs whereas a symmetric G mode at 1580 cm$^{-1}$ is typical of graphite. Furthermore, graphite also shows a D-band at 1350 cm$^{-1}$. Then, the unsymmetrical G mode and the low $I_D/I_G$ ratio observed in as sintered and deformed 3YTZP/SWCNT composites indicate the absence of severe damages in the CNTs structure after sintering and mechanical testing. In case of 3YTZP/MWCNTs composites, D and unsymmetrical G mode bands are observed, and the $I_D/I_G$ ratio in MWCNTs and in 3YTZP/MWCNTs composites prior to and after creep test are similar, pointing out that MWCNTs structure has not been damaged during sintering and mechanical testing. However, it is worth emphasizing that the $I_D/I_G$ ratio in 3YTZP/MWCNT is higher than in 3YTZP/SWCNT composites, revealing the lower quality of MWCNTs, which should contain a higher residual amount of graphite, compared to SWCNTs.

A creep curve for 3YTZP/MWCNT composite is shown in Figure 3. After each stress or temperature change, a brief transitory state followed by a steady state characterized by a constant strain rate $\dot{\varepsilon}$ is observed. Using the standard phenomenological creep equation (Eq. 1) the stress exponent $n$ and the activation energy $Q$ values have been calculated.

Figure 4 shows the log-log plot of the steady state strain rate versus stress obtained from all creep experiments at 1100 and 1200 ºC for monolithic 3YTZP, 3YTZP/SWCNTs and 3YTZP/MWCNTs composites. From the slopes of the linear regressions in Figure 4, average stress exponent n has been calculated in each case. Monolithic 3YTZP shows a stress exponent n=1.7 and n=2.1 at 1100 and 1200 ºC respectively. These values are close to 2, in agreement with results previously reported [22-24], where grain boundary sliding GBS accommodated by cationic diffusion throughout the lattice was recognized as the high temperature deformation mechanism. In case of composites, the stress exponent values are slightly higher but between 1.7 and 2.5. High temperature deformation mechanisms were studied in monolithic 3YTZP and 3YTZP/SWCNTs where mechanical results were analysed and discussed in the light of different models available in literature [24]. Among them, the generalized grain boundary sliding model accommodated by diffusional processes explained satisfactorily the high temperature mechanical behavior where grain boundary diffusion plays a fundamental role. In the framework of this model, it was argued that the incorporation of SWCNTs in a 3YTZP ceramic matrix influences yttrium segregation. Thus, grain boundary diffusion in composites is less hindered compared to monolithic 3YTZP and consequently a lower activation energy for 3YTZP/SWCNTs specimens compared to monolithic 3YTZP was obtained. In the case of 3YTZP/MWCNTs composites, the stress exponent $n$ and the activation energy values are similar to those obtained for 3YTZP/SWCNTs composites [24]. The absence of microstructural evolution in all deformed specimens, particularly the nonappearance of grain growth or changes in the shape factor for deformations even further than 20 % as shown in Table I, and stress exponent n values evidence that GBS is operating in monolithic 3YTZP, 3YTZP/SWCNTs and 3YTZP/MWCNTs composites.

It is worth emphasizing that monolithic 3YTZP exhibit the highest creep resistance, whereas 3YTZP/SWCNTs composites are the less creep resistant. Thus, for instance at 1200 ºC and 20 MPa monolithic 3YTZP exhibits a strain rate of about ~$10^{-7}$ s$^{-1}$, 3YTZP/MWCNTs composite deforms at ~$10^{-6}$ s$^{-1}$ and finally 3YTZP/SWCNTs composite at ~$10^{-5}$ s$^{-1}$. Then, there is a difference of about one order of magnitude in the strain rate between each specimen. At this point it should be noticed that differences in the grain size cannot justify such an increase in the strain rate, as can be calculated from equation (1) taking *p* between 1 and 3 [22] and grain size *d* values shown in Table I. So the reason for this different high temperature behaviour should be ascribed to the presence of CNTs, either SWCNTs or MWCNTs, surrounding 3YTZP grain boundaries. The lower creep resistance exhibited by nanocomposites compared to monolithic 3YTZP may attest the weak interfacial bonding between CNTs, either SWCNTs or MWCNTs, and 3YTZP ceramic grain boundaries. So the presence of both, SWCNTs or MWCNTs, makes easier grain boundary sliding during high temperature deformation and consequently composites exhibit a lower creep resistance compared to monolithic 3YTZP.

From our experimental results, the different creep behaviour exhibited by 3YTZP/SWCNTs and 3YTZP/MWCNTs composites is evident, since under same mechanical conditions there is a difference of about one order of magnitude in the strain rate. In this regard, it is worth emphasizing that mechanical properties of nanocomposites are CNTs distribution dependent. Strain rate, at 1200 ºC and 20 MPa, versus the surface density of CNTs agglomerates $\rho_s$ is plotted in Figure 5. In this figure, our experimental results and data taken from Ref [13] have been normalized to a grain size of 0.2 μm. Despite the complex relationship between the strain rate and $\rho_s$, it is possible to plot a phenomenological tendency between these two parameters using data from Ref [13], where a systematic study on the SWCNTs dispersion in 3YTZP and creep resistance was performed. Experimental results obtained in this work for 3YTZP/SWCNTs composite fits reasonably well to this tendency. Two limit situations can be imagined: (i) all SWCNTs are forming agglomerates instead of being located at the grain boundaries to facilitate GBS. So a mechanical behaviour similar to the monolithic zirconia can be expected. In fact, the strain rate of 3YTZP/SWCNTs composites decreases with the surface density of CNTs agglomerates. It is expected that for higher values of $\rho_s$ the strain rate of composites tend asymptotically to the strain rate value obtained for monolithic zirconia. (ii) No CNTs agglomerate is present in the material. In this perfect situation we could estimate the strain rate following the trend shown in figure 5.

3YTZP/MWCNTs composite shows a lower strain rate than that predicted from this tendency. This fact indicates that differences in the high temperature mechanical behavior between 3YTZP/SWCNTs and 3YTZP/MWCNTs composites should not be due only to a different level of CNTs dispersion in the 3YTZP ceramic matrix. Along with CNTs dispersion, there must be another parameter to justify such mechanical behavior. As described in section 2.1, both SWCNTs and MWCNTs followed a COOH-functionalization process through adsorption of negatively charged carboxyl group. So the chemical composition of the surface for both CNTs is qualitatively similar and then a similar interfacial bonding between CNTs, either SWCNTs or MWCNTs, and 3YTZP ceramic grain boundaries can be expected. Then, the main difference between SWCNTs and MWCNTs is related to the diameter and bundle length. COOH-functionalized MWCNTs have a bundle length between 1 and 5 μm whereas for SWCNTs is about three times shorter, as indicated in section 2.1. Since the average 3YTZP grain size is about 0.2 μm, each bundle of MWCNTs is surrounding and coupling the motion of a higher number of 3YTZP grains than in case of

SWCNTs. That could explain the lower creep rate exhibited by 3YTZP/MWCNTs composite compared to that predicted by the trend shown in figure 5. Finally, the low friction exhibited by MWCNTs for telescopic extension [25] could in principle favour GBS, however the high length and level of interwoven of MWCNTs between 3YTZP grains seem to hinder the sliding of the inner shells in the MWCNTs and consequently it has no important influence on GBS.

## 4. Conclusions

Creep behavior in monolithic 3YTZP and 3YTZP containing 2.5 vol% of either single-walled carbon nanotubes (SWCNTs) or multi-walled carbon nanotubes (MWCNTs) has been investigated in this work. Microstructural studies confirm the absence of severe damages in CNTs after sintering and testing. Experimental results show that the incorporation of CNTs, either SWCNTs or MWCNTs, into a 3YTZP matrix makes the composite less creep resistant with respect to monolithic zirconia. Mechanical results point out GBS as the deformation mechanism responsible of the high temperature mechanical behavior in all specimens, where CNTs make easier GBS during high temperature deformation. However, as discussed in the paper, creep behavior is not influenced by the type of CNTs, either SWCNTs or MWCNTs, but rather by the bundle length and the level of dispersion of CNTs in the 3YTZP ceramic matrix.

**Acknowledgements**

This work was financially supported by the European Regional Development Fund and the Spanish "Ministerio de Economía y Competitividad" through the projects MAT2009-11078,

MAT2012-34217 and the project from the Andalucia Government P12-FQM-1079. M. C-R thanks the JAE-doc contract awarded by the Spanish CSIC, co-financed by the European Social Fund.

**FIGURE CAPTIONS**

Fig. 1. HRSEM micrographs of a) polished and thermally etched surface of monolithic 3YTZP, and fracture surfaces of b) 3YTZP/SWCNTs as-sintered composite, c) 3YTZP/SWCNTs composite after creep test, d) 3YTZP/MWCNTs as-sintered composite and e) 3YTZP/MWCNTs composite after creep test.

Fig. 2. Raman spectra of a) SWCNTs, monolithic 3YTZP and 3YTZP/SWCNTs and b) MWCNTs, monolithic 3YTZP and 3YTZP/MWCNTs. For composites, Raman spectra of specimen prior and after creep test have been included.

Fig. 3. Creep curve, showing the stress and temperature changes to obtain the stress exponent n and the activation energy Q for 3YTZP/MWCNTs nanocomposites.

Fig. 4. Strain rate-stress logarithm plot for monolithic 3YTZP (circles), 3YTZP/SWCNTs (triangles) and 3YTZP/MWCNTs (squares) composites. Filled and empty markers correspond to 1200 and 1100 ºC, respectively. The average stress exponents n are also shown.

Fig. 5. Strain rate logarithm, at 20 MPa and 1200 ºC, versus surface density of CNTs agglomerates plot for composites tested in this work and from Ref [13]. Data have been normalized to a grain size of 0.2 μm. The strain rate of the monolithic 3YTZP is also shown for the sake of comparison.

**TABLES**

Table I. Morphological parameters (average grain or agglomerate size *d* and shape factor *F*) for 3YTZP, 3YTZP/SWCNTs and 3YTZP/MWCNTs, as-sintered and after creep tests. Relative density $\rho_r$ and surface density $\rho_s$ of CNTs agglomerates are also shown.

| Material | 3YTZP grains | | CNTs agglomerates | | | $\rho_r \pm 0.5$ (%) |
|---|---|---|---|---|---|---|
| | *d* (μm) | *F* | $\rho_s$ (%) | *d* (μm) | *F* | |
| 3YTZP | 0.27 ± 0.10 | 0.72 ± 0.07 | | | | 99.5 |
| 3YTZP deformed | 0.25 ± 0.12 | 0.74 ± 0.07 | | | | |
| 3YTZP/SWCNTs | 0.20 ± 0.09 | 0.71 ± 0.09 | 0.3 ± 0.1 | 0.5 ± 0.3 | 0.83 ± 0.15 | 99.7 |
| 3YTZP/SWCNTs deformed | 0.20 ± 0.09 | 0.71 ± 0.09 | | | | |
| 3YTZP/MWCNTs | 0.23 ± 0.09 | 0.69 ± 0.08 | 0.7 ± 0.1 | 0.6 ± 0.1 | 0.66 ± 0.18 | 99.4 |
| 3YTZP/MWCNTs deformed | 0.22 ± 0.10 | 0.70 ± 0.13 | | | | |

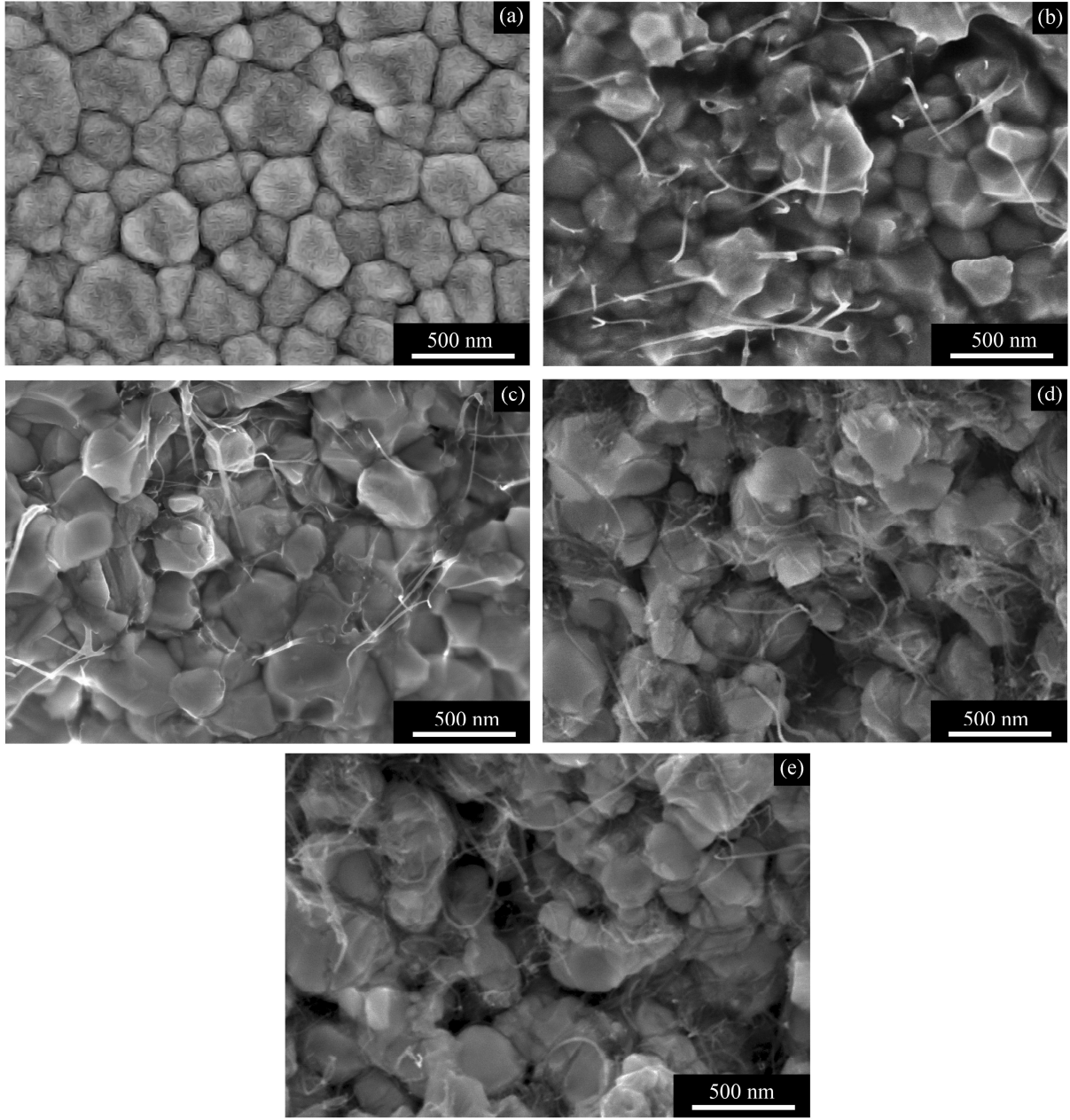

Fig. 1

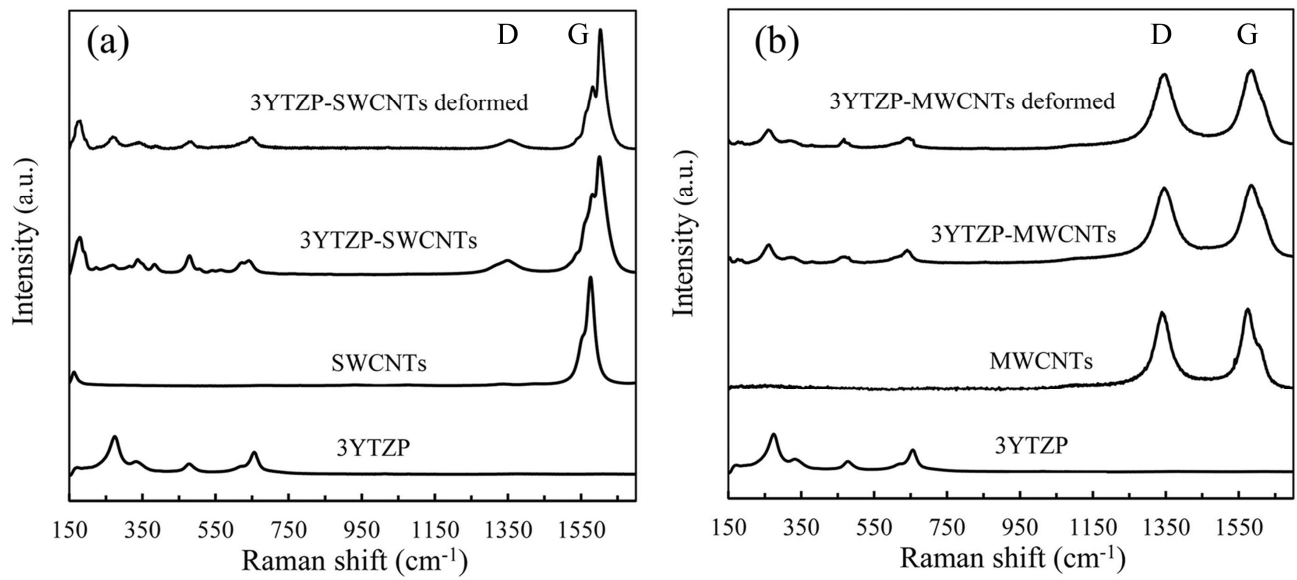

Fig. 2

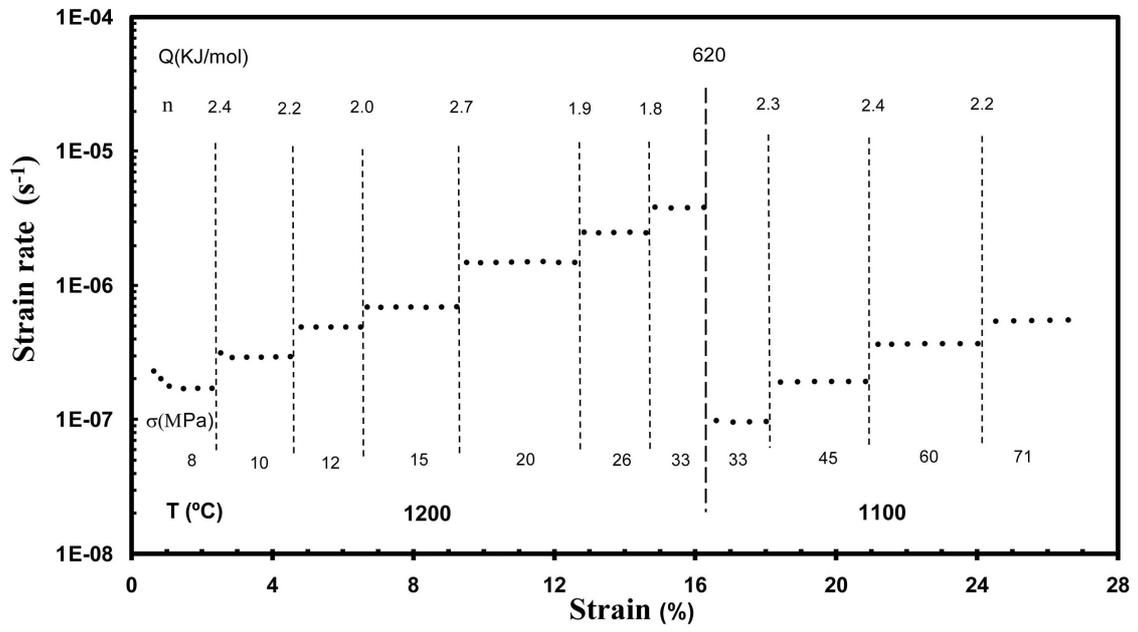

Fig. 3

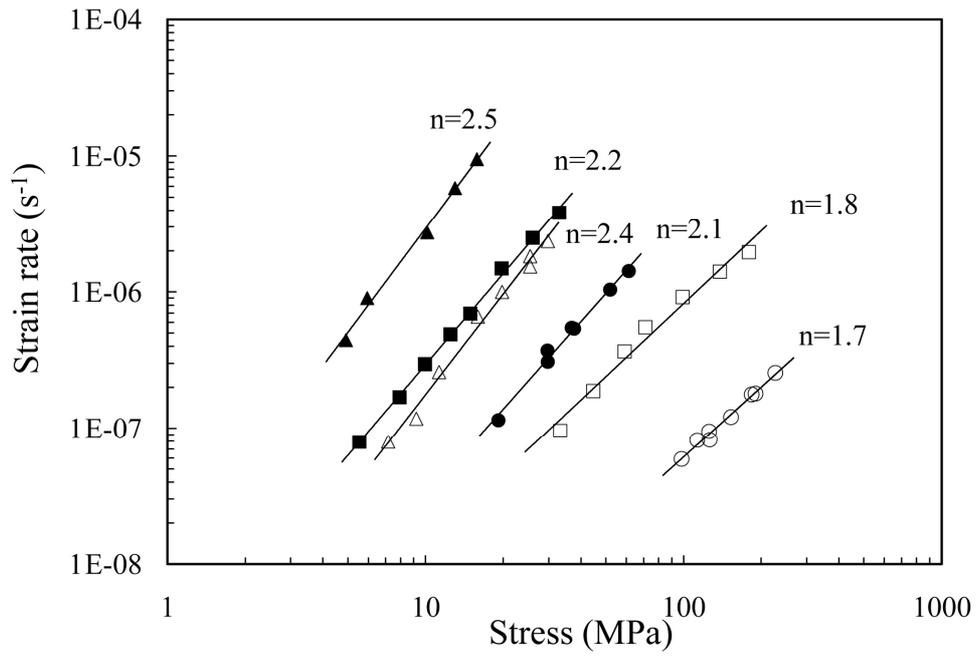

Fig. 4

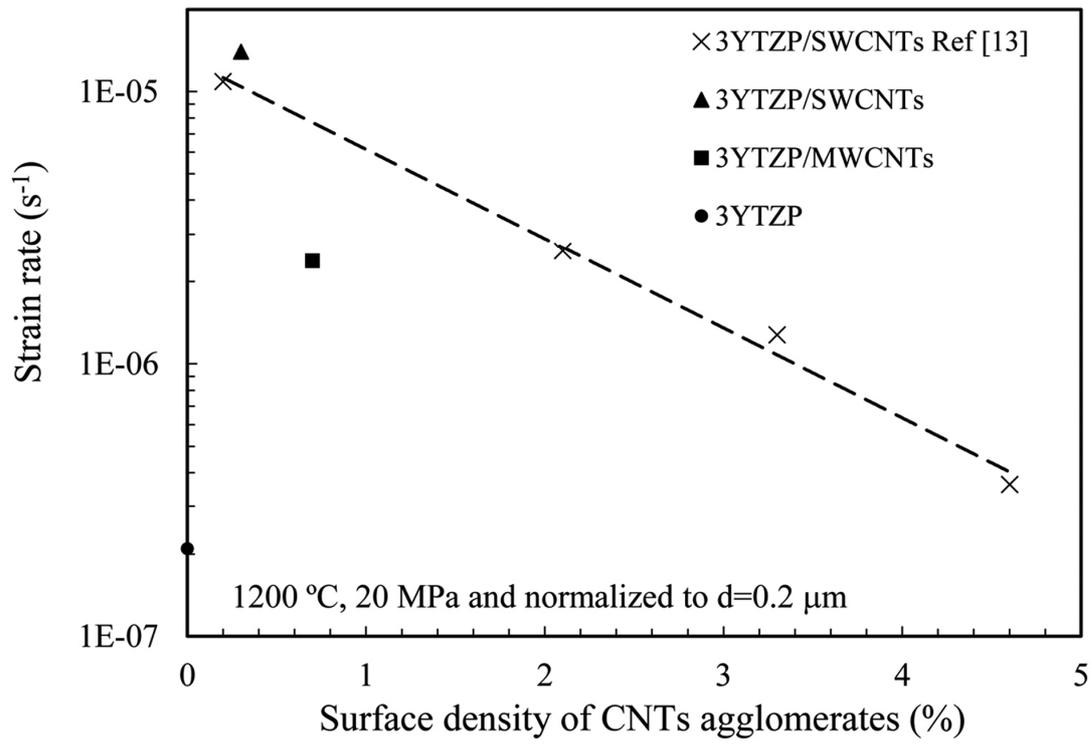

Fig. 5